\documentclass[preprint,aps,superscriptaddress]{revtex4}
\usepackage{graphicx,epsfig,amssymb,amsmath,array}
\usepackage{relsize,placeins,float}
\usepackage[usenames]{color}

\begin{document}
\title{Fractional damping effects on the transient dynamics of the Duffing oscillator}
\author{Mattia Coccolo}
\affiliation{Nonlinear Dynamics, Chaos and Complex Systems Group, Departamento de F\'{i}sica,
Universidad Rey Juan Carlos, Tulip\'{a}n s/n, 28933 M\'{o}stoles, Madrid, Spain}
\author{Jes\'{u}s M. Seoane}
\affiliation{Nonlinear Dynamics, Chaos and Complex Systems Group, Departamento de F\'{i}sica,
Universidad Rey Juan Carlos, Tulip\'{a}n s/n, 28933 M\'{o}stoles, Madrid, Spain}
\author{Stefano Lenci}
\affiliation{Department of Civil and Building Engineering, and Architecture, Polytechnic University of Marche, 60131 Ancona, Italy}
\author{Miguel A.F. Sanju\'{a}n}
\affiliation{Nonlinear Dynamics, Chaos and Complex Systems Group, Departamento de F\'{i}sica,
Universidad Rey Juan Carlos, Tulip\'{a}n s/n, 28933 M\'{o}stoles, Madrid, Spain}
\affiliation{Department of Applied Informatics, Kaunas University of Technology \\ Studentu 50-415, Kaunas LT-51368, Lithuania}
\date{\today}

\begin{abstract}
We consider the nonlinear Duffing oscillator in presence of fractional damping which is characteristic in different physical situations. The system is studied with a smaller and larger damping parameter value, that we call the underdamped and overdamped regimes. In both we have studied the relation between the fractional parameter, the amplitude of the oscillations and the times to reach the asymptotic behavior, called {\it asymptotic times}. In the overdamped regime, the study shows that, also here, there are oscillations for fractional order derivatives and their amplitudes and asymptotic times can suddenly change for small variations of the fractional parameter. In addition, in this latter regime, a resonant-like behavior can take place for suitable values of the parameters of the system. These results are corroborated by calculating the corresponding $Q-$factor. We expect that these results can be useful for a better understanding of fractional dynamics and its possible applications as in modeling different kind of materials that normally need complicated damping terms.

\end{abstract}

\maketitle

\section{Introduction}\label{sec:introduction}

A dynamical system spends a certain amount of time before arriving asymptotically to a steady state. This can be a fixed point, a limit cycle of different periods, quasi-periodic or a chaotic attractor. Usually, the study of dynamical systems is focused once the system has arrived to their steady state, which is enough when the possible relevance of the transient dynamics can be omitted, which however is not always the case. Actually, the duration of the transient dynamics can vary from very short times to very long times. Furthermore, one interesting phenomenon that can be found is transient chaos. The rich dynamical behavior of this kind of transient phenomena explains why they have been studied so much in the past few years. In fact, transient phenomena and their study has been carried out by several branches in science, like ecological systems \cite{Hasting2004,Hasting2018,Morozov2020}, neuroscience \cite{Rabinovich2006, Rabinovich2008} gravitational waves \cite{Thrane},  mathematics \cite{Morozov2016} and meteorology \cite{Cantisan}, among others. All these previous works have been developed considering ordinary differential equations.

In studies about materials, like materials with memory \cite{Dafermos}, or vibratory structures \cite{Lu}, there might appear complex forms of damping terms \cite{Chellaboina,De,Fabrizio,Casciati,Elliott} to model their transient behaviors, though still using ordinary differential equations. On the other hand, the use of fractional derivatives to model memory effects in materials has been commonly used. This implies that its state does not depend just on the current time and position but also on the previous states. Some examples of this behavior are: an electrical circuit component whose resistance depends on all the charge that has passed through it over a fixed length of time, viscoelastic materials or vibratory structures, among others. Systems with memory effects can be very difficult to model and analyze with ordinary differential equations. However, by using fractional derivatives the memory effects can be easily incorporated. Therefore, fractional calculus could prove to be a very useful tool for analyzing systems with complex transient behaviors, normally modeled with complicated damping terms \cite{Horr}.

As a consequence of the previous considerations, we have decided to study the relationship between the fractional derivative terms, the amplitude reached by the oscillations in the steady state and the time it takes to reach its final asymptotic behavior. Also, we study that relationship in a case of smaller damping, $\mu=0.15$, and higher damping, $\mu=0.8$.  For our study, we have used the well-known Duffing oscillator with a fractional derivative instead of the first-order derivative to model the damping term. In the non-fractional case, the Duffing oscillator, when $\mu=0.15$, can present a very rich variety of dynamical behaviors such as chaos, periodicity, etc. Here, the influence of the damping term is not so relevant and therefore we consider the system to be in the underdamped regime.  On the other hand, when $\mu=0.8$, the dynamics are driven by the dissipation term and almost all the trajectories fall into a fixed point. Therefore, the dynamical behavior is quite trivial and we consider the system to be in the overdamped regime. Previous researches on the Duffing system with a fractional damping term \cite{Boroviec,Sheu} were focused on the influence of the order of the derivative or the amplitude of the excitation on the system dynamics. Its sensitivity to the initial conditions has been also studied \cite{Syta}, as well as the effect of using anharmonic external perturbations looking for geometrical resonances \cite{SJimenez}. The stability analysis has been also carried out \cite{Haeri,Petras} in some particular cases.

To summarize, the main goal of this work is to study the role of the fractional derivative parameter on the amplitude of the system oscillations and to analyze its influence on the time needed for the system to reach its steady state, that we call {\it the asymptotic time}. Then, by studying the oscillation amplitudes we demonstrate that a resonance-like phenomenon can be triggered for high values of dissipation and for certain forcing amplitudes and fractional derivative orders. Finally, we want to illustrate that the complicated damping term used to study the transient behaviors of different kind of materials can be substituted by a fractional derivative term.

The organization of this paper is as follows. In Sec.~\ref{sec:2}, we describe the Duffing oscillator with a fractional order damping. In Sec.~\ref{sec:3}, we compute the amplitude of the oscillations as a function of the fractional parameter $\alpha$ in the underdamped regime. In Sec.~\ref{sec:4}, a study of the time to reach the asymptotic behavior is carried out. Furthermore, we extend the previous results for the overdamped case in Sec.~\ref{sec:5}. Finally, Sec.~\ref{sec:conclusions} summarizes the main findings and possible applications of this work.

\section{model description}\label{sec:2}
\begin{figure}[htbp]
  \centering
   \includegraphics[width=10.0cm,clip=true]{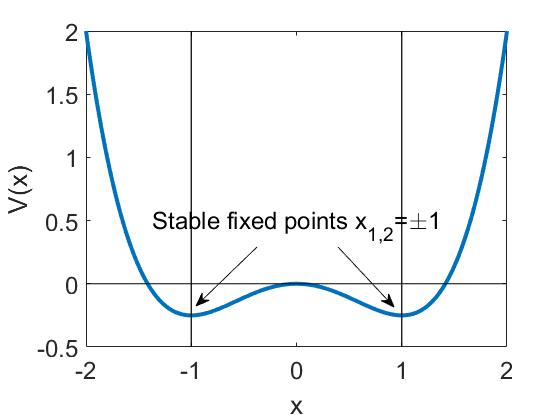}
   \caption{The double-well potential of the Duffing oscillator indicating the stable fixed points. }
\label{fig:1}
\end{figure}

\begin{figure}[htbp]
  \centering
   \includegraphics[width=16.0cm,clip=true]{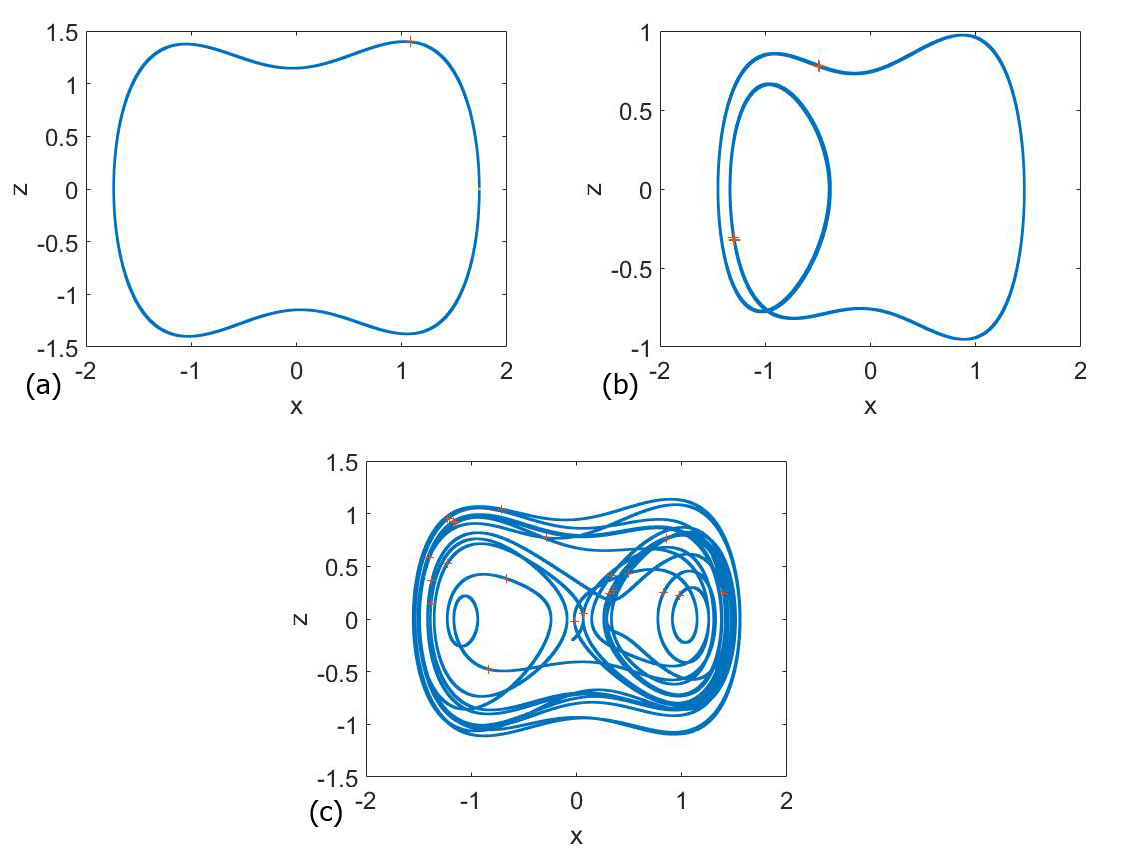}
   \caption{Panels (a-c) show the orbits of the system for $\mu=0.15$, $F=0.3$, $\omega=1$ with $\alpha =0.6~(a)$, $\alpha=0.8$~(b) and $\alpha=1$~(c), respectively. The initial condition is fixed as $(x_0,y_0,z_0)=(0.2,0.3,0)$. All the orbits are drawn discarding the transient behaviors, that are different from each case.}
\label{fig:2}
\end{figure}

As we have already mentioned, a Duffing oscillator with a fractional derivative term instead of the first derivative, i.e., $\dot{x}(t) \to D^{\alpha}x(t)=d^{\alpha}x(t)/dt^{\alpha}$ as a damping term, will be considered.  Then, the fractional Duffing oscillator with a fractional damping reads
 \begin{equation}
     \frac{d^2x}{dt^2}+\mu\frac{d^{\alpha}x}{dt^{\alpha}}-x+x^3=F\cos{\omega t},
 \end{equation}
 where $m=1$,
 $\mu$ is the damping coefficient, $F$ and $\omega$ denote both the amplitude and frequency of the external forcing, respectively, and $\alpha$ is the fractional parameter. The model describes the dynamics of a unit mass in a double-well potential, as shown in Fig.~\ref{fig:1}, and eventually exhibits chaotic behavior for a particular choice of parameters. This system, in absence of dissipation and forcing, presents two fixed points located at the bottom of every well at $x_{1,2}=\pm 1$. In order to integrate the system, we use the {\it Gr\"{u}nwald-Letnikov} fractional derivative, as in a previous work \cite{Coccolo_fr}, in which it has already been tested for the fractional Helmholtz oscillator with satisfactory results. Consequently, the system is composed by a set of three fractional differential equations that yield

\begin{equation}\label{model}
\left\lbrace\begin{array}{l}
 D^{\alpha}x=y, \medskip\\
 D^{1-\alpha}y=z, \medskip\\
 Dz= \mu y + F\cos{(\omega t)}+x-x^3,
\end{array}\right.
\end{equation}
where $z$ is an auxiliary component coming from the transformation of the model into a fractional order system. The integration step is $\pi/2000$, the maximum time is $t=600$ in the underdamped case ($\mu=0.15$) and $t=300$ in the overdamped case ($\mu=0.8$), values that in general, are large enough for the system to reach the steady state.
Also, the integrator has been checked out comparing some preliminary results with \cite{Syta}, as shown in Figs.~\ref{fig:2}(a-c).  So that, in the next section we can start discussing our results with confidence. Moreover, those orbits show the basic dynamics of this oscillator for different values of the fractional parameter $\alpha$.

\section{Oscillations amplitude in the underdamped case}\label{sec:3}
Here, we start our research for the underdamped case taking as parameter values $\mu=0.15$ and $\omega=1$. For the numerical analysis we have used a MATLAB in-house code generated specifically for this purpose. We have carried out numerical simulations with different initial conditions and we have seen that, although the results are numerically different, the general trends of the curves are qualitatively similar. So we decided to carry on the study with the initial condition $(x_{0}, y_{0}, z_{0}) = (0.2, 0.3, 0.0)$, as in Ref.~\cite{Syta}, and we expect that our conclusions are of general validity and not specific for the considered boundary conditions.
Notice that all figures shown here are about the oscillations in the $x$ coordinate, because showing the same figures for the $z$ coordinate do not provide any further relevant information.

\begin{figure}[htbp]
  \centering
   \includegraphics[width=16.0cm,clip=true]{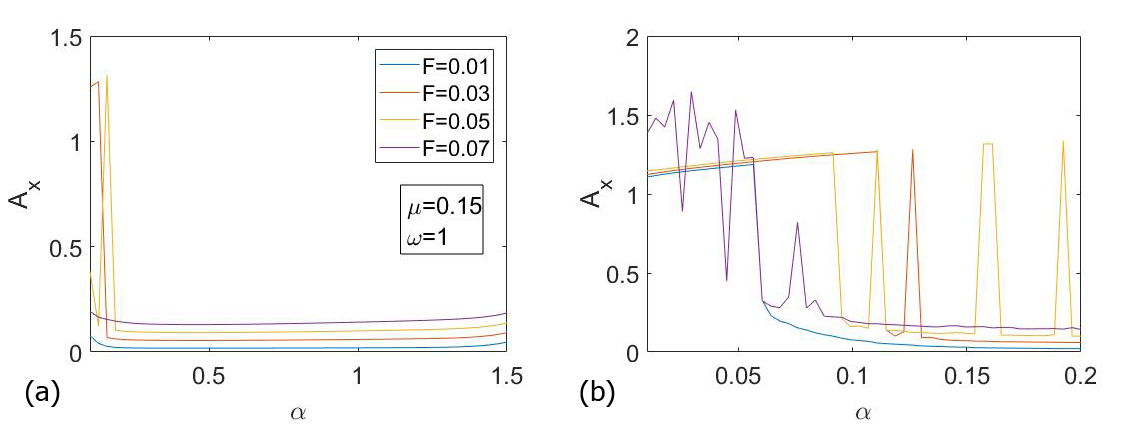}
   \caption{Panel~(a) shows the amplitude of the oscillations in function of $\alpha$ for different forcing amplitudes $F=0.01$, $F=0.03$, $F=0.05$ and $F=0.07$, $\omega=1$ and $\mu=0.15$. We can identify the region of interest in the left part of the figure. Panel~(b) is a zoom of panel~(a) in the region of interest $0.01<\alpha<0.2$, for the same parameters values as panel~(a), where clear variations of $A_x$ can be seen.}
\label{fig:3}
\end{figure}

In Fig.~\ref{fig:3}, we show the amplitudes of the $x$ variable oscillations $A_x$ in function of the fractional parameter $\alpha$ for different values of the forcing amplitude $F$. The simulations have been stopped when $F=0.07$, since for higher values of $F$ the trend is similar.  Observe that, when the forcing amplitude is higher than $F=0.07$, its effects on the dynamical behavior of the system are dominant. This fact happens since, for these values of the forcing amplitudes, small values of $\mu$ does not allow the damping to play a relevant role, and the fractional parameter $\alpha$, alone, is not able to change this behaviour. However, for all the $F$ values we can observe the flat amplitude line in Fig.~\ref{fig:3}(a) for $\alpha>0.2$, which is not interesting for our purposes, since the oscillations amplitudes do not change significantly in function of $\alpha$. In fact, the higher the forcing amplitude, the higher its influence on the dynamics of the system and the smaller the impact of the fractional derivative.
On the contrary, for smaller $\alpha$ values sudden changes of the amplitude $A_x$ occur and a region of interest shows up.  That region is presented in Fig.~\ref{fig:3}(b) where we show the same plot for the same forcing values $F$, but for $\alpha$ values between $0.01$ and $0.2$. Here, it is possible to realize that for $F=0.01$, after an initial plateau of higher amplitudes, the amplitudes decrease and remain constant for the rest of values of $\alpha$. The plateau can be related with the limited influence of the damping term, due to the small derivative order, that makes the term almost negligible. Then, for $F=0.03$ and $F=0.05$,  after the initial plateau, some peaks appear for higher $\alpha$ values. On the other hand, for $F=0.07$ the plateau disappears, giving birth to some peaks and then the oscillations amplitude decreases to a constant value.

\begin{figure}[htbp]
  \centering
   \includegraphics[width=15cm,clip=true]{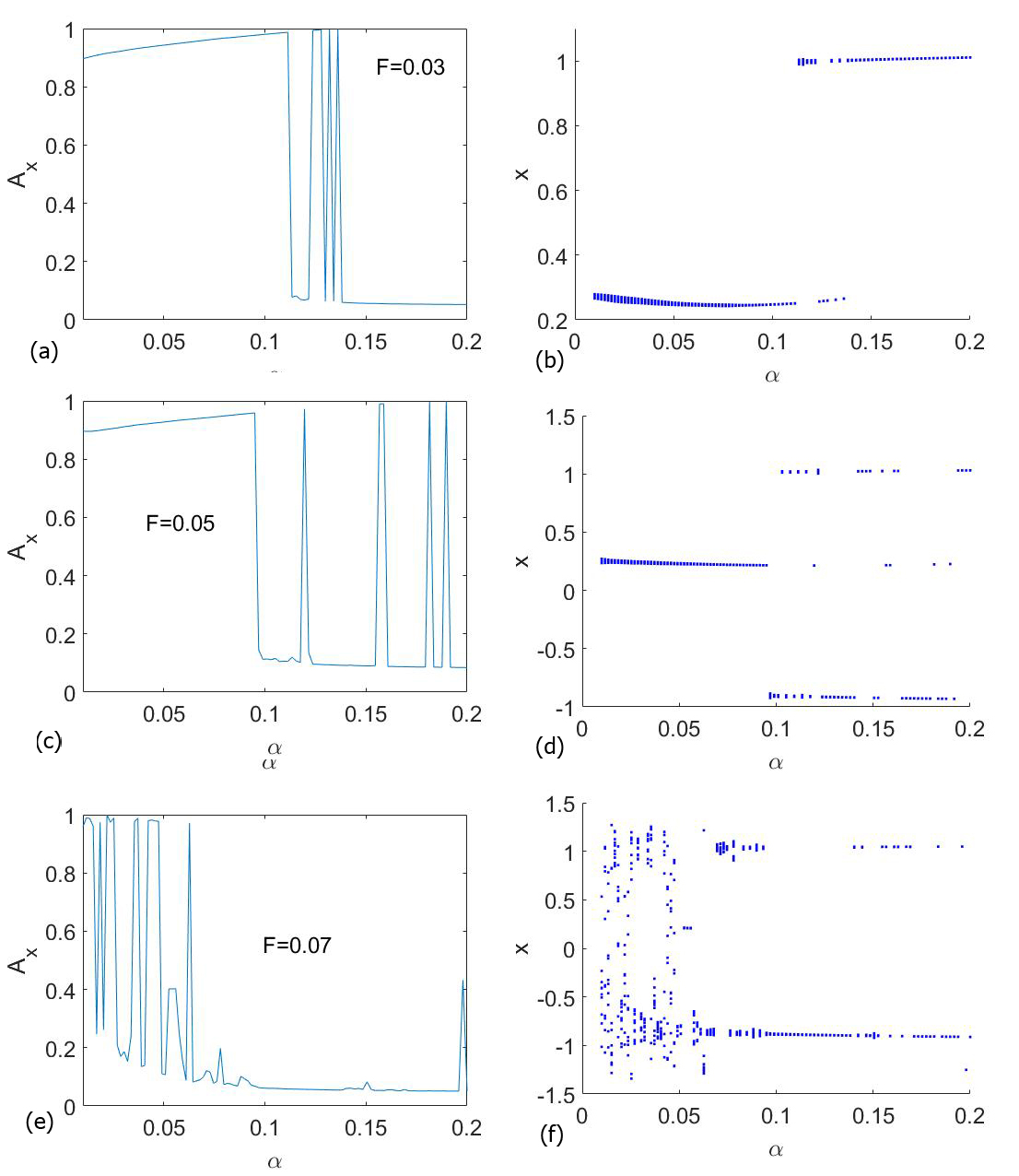}
   \caption{Panels (a), (c) and (e) show the amplitude of the oscillations for  $F=0.03,~F=0.05$ and $F=0.07$ respectively and $\mu=0.15$ with higher resolution than in Fig.~\ref{fig:3}(b). Those figures are compared with a diagram of the asymptotic behaviors (panels~(b), (d) and (f)), i.e.,  they show the set of points of the stroboscopic map for the steady state. By comparing panels (a) and (b) with panels (c) and (d), it is possible to see that the peaks are related with a change in the dynamics of the system.  Conversely, by comparing panels (e) and (f)  it is possible to see that the peaks are related with a chaotic dynamics of the system.}
\label{fig:4}
\end{figure}

In order to study the dynamics beneath these changes in the amplitudes, we have plotted the oscillations amplitude in function of $\alpha$ for the same $F$ values but with a higher resolution in Figs.~\ref{fig:4}. In addition, for all the cases in this last figure we sketch a diagram of the asymptotic behavior as a function of the $\alpha$ parameter, in order to have a better understanding of the dynamics that generates the changes in the amplitude. In these diagrams, we depict the set of points of the stroboscopic map related with the steady state of the system oscillations. This means that the points in the asymptotic behaviors diagrams just display the dynamics in the steady state, they are not related with the oscillations amplitude.

In Fig.~\ref{fig:4}(a) it is possible to recognize some other peaks than the ones in Fig.~\ref{fig:3}(b), coming out close to each other. Furthermore, by comparing it with Fig.~\ref{fig:4}(b), we can see that the end of the plateau and the start of the first peak overlap with a drastic change in the asymptotic behaviors, so as the other peaks. A similar pattern is shown in Fig.~\ref{fig:4}(c) and Fig.~\ref{fig:4}(d).

\begin{figure}[htbp]
  \centering
   \includegraphics[width=16.0cm,clip=true]{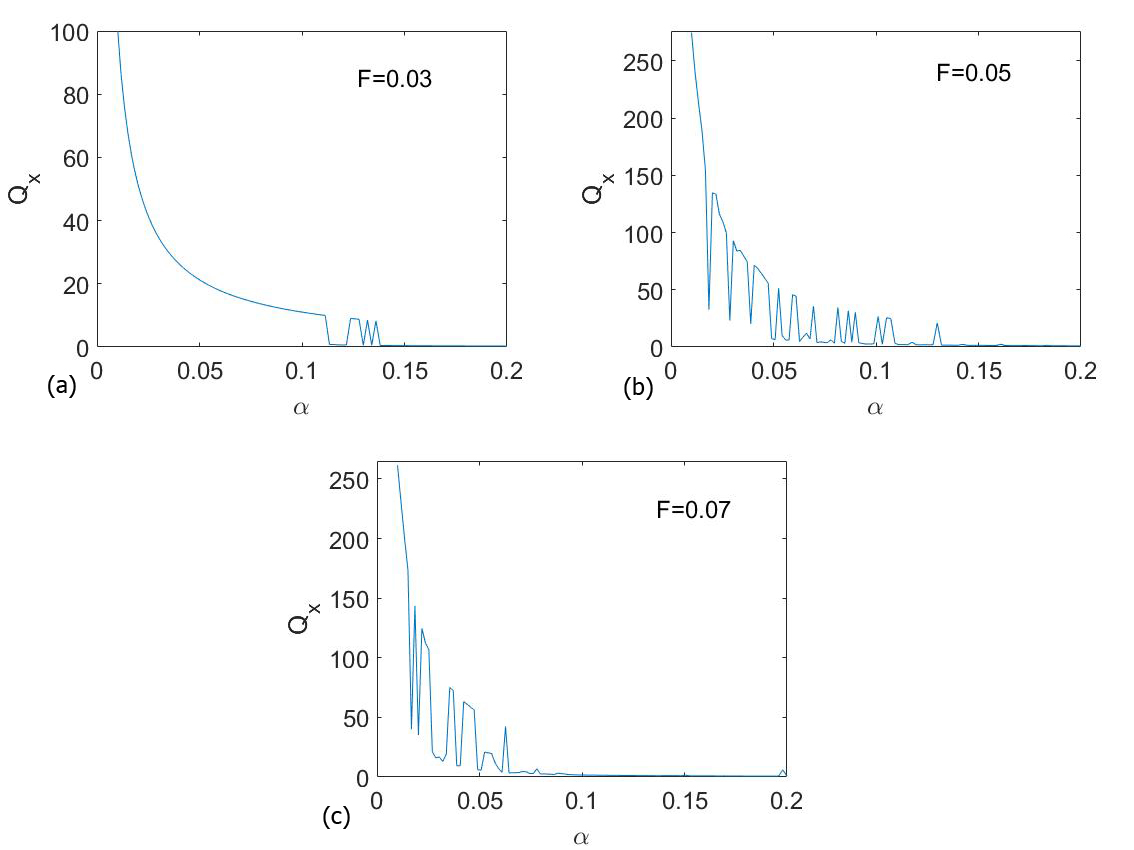}
   \caption{These panels show the $Q-$factor of the Figs.~\ref{fig:4}(a),~\ref{fig:4}(c) and~\ref{fig:4}(e). It is possible to see that some peaks show up, although they are related with changes in the dynamics of the system, rather than with the appearance of a possible resonance.}
\label{fig:5}
\end{figure}

On the contrary, Figs.~\ref{fig:4}(e) and~\ref{fig:4}(f)  show us that the disappearance of the plateau for a set of peaks is related with the emergence of a chaotic region, that is why the oscillation amplitudes (as the maximum in time of $x(t)$) oscillate  and stop a reliable characterization of the oscillations. This region starts to appear for $\alpha \simeq 0.06$.  There is also a smaller peak for $\alpha=0.2$, but it is alone and no other shows up for further values. As explained before, for even higher values of $F$ the pattern is similar to this last one, so that they are of no interest in our exploration. As we can see, except in the chaotic region of Fig.~\ref{fig:4}(f), the amplitude of the peaks in the previous figures does not reach the two wells of the potential. In fact, in Fig.~\ref{fig:4}(d) the points of the asymptotic behaviors diagram jump from one solution on the top to another to the bottom or to the center. In addition, in Fig.~\ref{fig:5} we have plotted for each case the $Q-$factor, providing an idea of how much the signal is amplified by the $\alpha$ parameter. This is possible by calculating the sine and cosine components:
\begin{equation}\label{eq:3}
    B_s=\frac{2}{nT}\int^{nT}_0{x(t)\sin{\omega t}dt}\quad\text{and}\quad B_c=\frac{2}{nT}\int^{nT}_0{x(t)\cos{\omega t}dt}
\end{equation}
where $T=2\pi/\omega$ and $n$ is an integer. Then we can find the dependence on $\alpha$ with $Q=\sqrt{B^2_s+B^2_c}/\alpha$. This shows that several peaks still show up. Although, as stated before, they are more related with the changes on the dynamics rather than with the appearance of a resonance phenomenon.

\FloatBarrier

 \begin{figure}[htbp]
  \centering
   \includegraphics[width=16.0cm,clip=true]{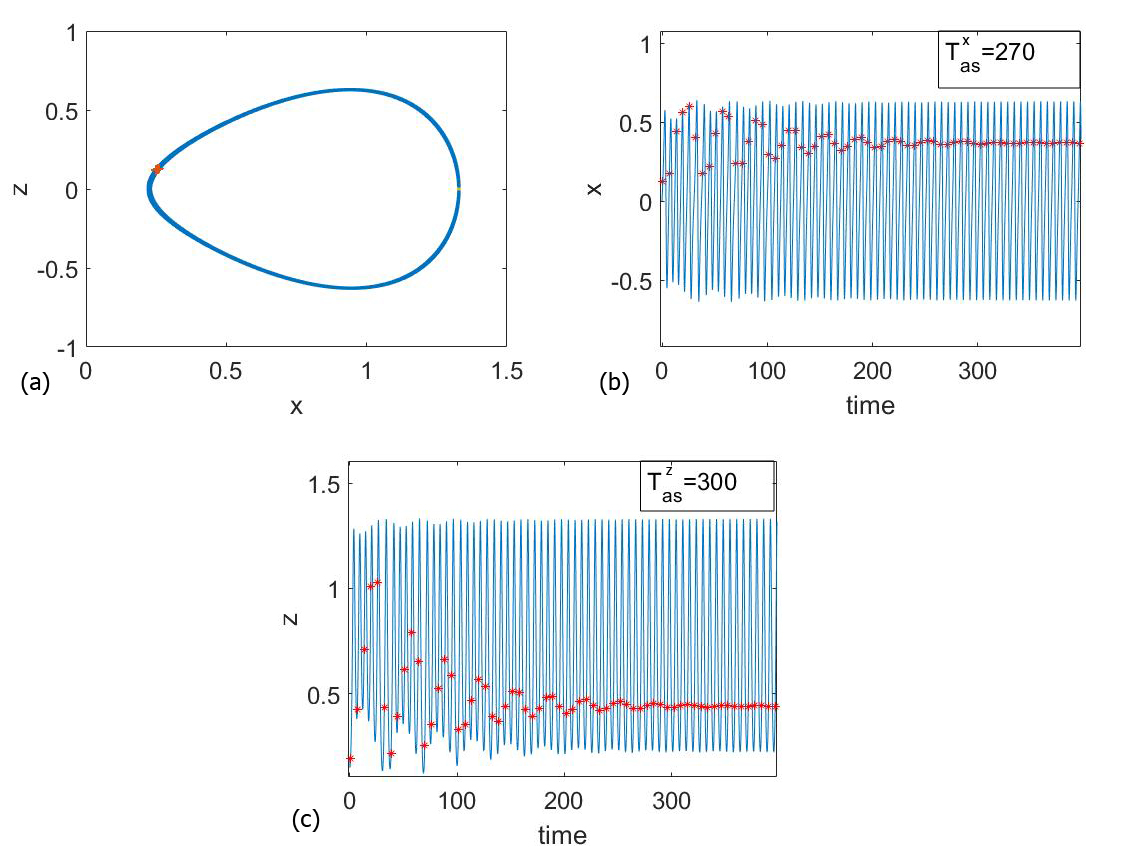}
   \caption{The panel on the top left (a) shows the phase space for $F=0.03,~\alpha=0.11,~\omega=1$ and $\mu=0.15$. Then, the time series of the $x$ coordinate (panel~(b)) and  the time series of the $z$ coordinate (panel~(c)), of an orbit with $F=0.03,~\alpha=0.11,~\omega=1$ and $\mu=0.15$ are also plotted. Here, it is possible to appreciate the asymptotic time ($T_{as}$) given by the start of the alignment of the points of the stroboscopic map in the time series.}
\label{fig:6}
\end{figure}

\begin{figure}[htbp]
  \centering
   \includegraphics[width=16.0cm,clip=true]{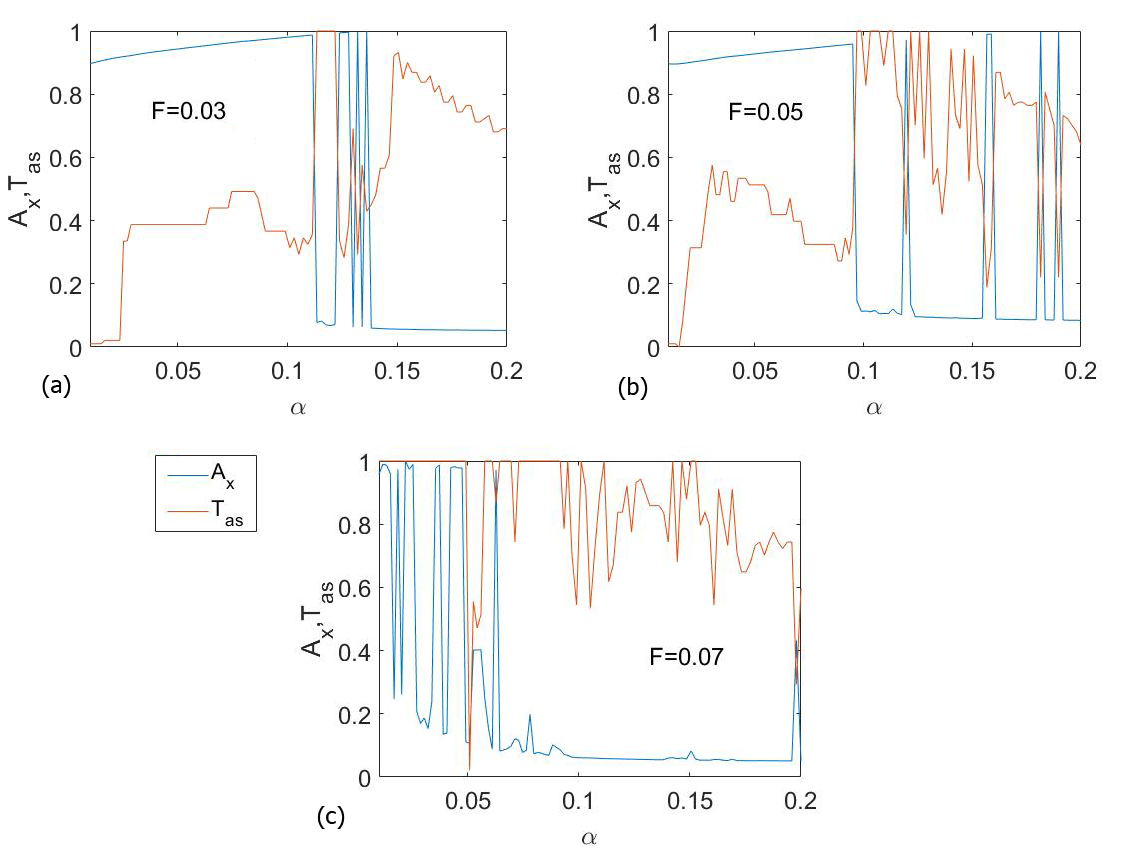}
  \caption{ The asymptotic times $T_{as}$ overlapped with the amplitude plots of Figs.~\ref{fig:4} are depicted. In all the panels, the amplitudes and the asymptotic times are normalized to one, to be able to show them in the same plot. The maximum of the asymptotic time is $t=600$. In all panels, we can observe that the asymptotic times change abruptly when the $\alpha$ parameter changes. Panels (a) and (b) show that the lowest asymptotic times are related with the amplitude peaks. Panel~(c) shows a different behavior, due to the chaotic region displayed in Fig.~\ref{fig:4}(f).}
\label{fig:7}
\end{figure}

\section{Asymptotic time in the underdamped case}\label{sec:4}

Now, we reach the core of our research work by studying how the parameter $\alpha$ can influence, not only the oscillations amplitudes, but also the time for which the oscillator reaches its asymptotic behavior that we have called {\it the asymptotic time}. This asymptotic time $T_{as}$ has been calculated through the estimation of the alignment of the points in a stroboscopic map, as shown in Fig.~\ref{fig:6}. For the chaotic solutions, as there is no alignment of the stroboscopic map points, the $T_{as}$ would be the maximum integration time. Recall that in the underdamped case $t=600$.
In this figure, the asymptotic times have been plotted for $F=0.03,~\alpha=0.11,~\omega=1$ and $\mu=0.15$. Here, for a better visualization, the maximum time of integration has been set at $t=400$. In Fig.~\ref{fig:6}(c), we have plotted the asymptotic time for the $x$ and $z$ coordinate ($T^x_{as},T^z_{as}$), where the asymptotic time difference between them is comparable with the sampling of the stroboscopic map. In other words, the asymptotic times of the two variables, $x$ and $z$, differ just by $(T^x_{as}-T^z_{as})/P=4.77$ periods of the system oscillations, with $P=2\pi/\omega=6.28$ the period.

Then, in Fig.~\ref{fig:7}, we show plots of both the oscillations amplitudes (blue lines) and the asymptotic time (orange lines) as a function of the $\alpha$ parameter for different values of the forcing amplitude. These figures, for the sake of clarity, have been normalized to a maximum value of one, in order to show in the same figure both the amplitude reached and the asymptotic times. The real maximum of the amplitudes is around $1$, as shown in Figs.~\ref{fig:4}(a),~\ref{fig:4}(c) and~\ref{fig:4}(e) and for the asymptotic times is $T_{as}=600$.

We can notice that in Figs.~\ref{fig:7}(a) and~\ref{fig:7}(b), whenever a peak of the amplitude shows up the asymptotic time decreases, while we have a maximum in the asymptotic time when the amplitude reaches a minimum.  In Fig.~\ref{fig:7}(c) the maximum of the asymptotic time is related with the chaotic behavior shown in Fig.~\ref{fig:4}(e). In all the figures the asymptotic times can reach their maximum also when the dynamics is not chaotic. This means that the oscillations need a time longer than $t=600$ to asymptotically fall to the attractor. As an example, we can take  the case of Fig.~\ref{fig:7}(a) for $0.11<\alpha<0.122$, where the amplitude is minimal and the asymptotic times are maximum. If we compare that figure with Fig.~\ref{fig:4}(b), we can realize that the asymptotic behavior of the system dynamics in that region is to fall to the fixed point of the associated Poincar\'e map, $x_1=1$. So, the oscillations were still settling to the fixed point at $t=600$.

For the sake of our goal, we want to underline that the asymptotic times change depending on the $\alpha$ value, but they jump from a value to another without following a discernible trend. This is a first indicator that it could be possible to model complicated forms of the damping terms in the study of some materials with a fractional derivative term. Although, for what is shown in these last figures the variations of the asymptotic times are difficult to predict.
 \begin{figure}[htbp]
  \centering
   \includegraphics[width=16.0cm,clip=true]{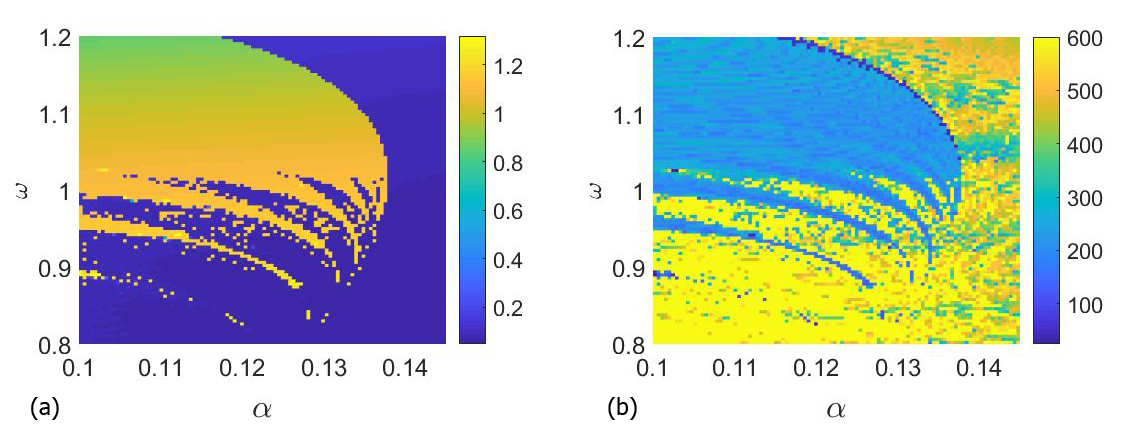}
   \caption{The left panel (a) shows, for $F=0.03$, the amplitude parametric set.  The right panel (b) shows the asymptotic time parametric set. In both cases the parameters used are $\alpha$ and $\omega$. We can recognize a region of parameter values for which we have high oscillations amplitudes and another one for which we have low oscillations amplitudes. Moreover, it is possible to see that one figure is the negative of the other (the highest oscillations amplitudes in the left figure are related with the lowest asymptotic times in the right figure) and that there are regions in which the colors are intermingled, showing a clear fractalization of the boundaries. This means that the oscillation amplitude and the asymptotic times for parameter values placed on the regions boundaries are unpredictable.}
\label{fig:8}
\end{figure}
\FloatBarrier

The relation of the oscillations amplitude and the asymptotic times and their dynamics can be studied changing both the fractional parameter $\alpha$ and the frequency $\omega$ values in a parameter set. This is a figure in which we vary the values of two parameters to show the variation of a magnitude with a color code indicated by the bar on the right of the figure. In fact, Figs.~\ref{fig:8}~(a) and (b) show that kind of plot for a value of the forcing amplitude $F=0.03$, changing the parameters $\omega$ and $\alpha$.  The range of $\alpha$ and $\omega$ are $0.1\leq\alpha\leq0.145$ and  for $0.8\leq\omega\leq1.2$. The frequency width has been decided in order to extend our understanding of the fractional parameter role on the dynamics of the system also for frequency values near $\omega=1$.  In Figs.~\ref{fig:8}~(a) and (b), it can be observed that the amplitude and the asymptotic time are one the negative of the other, i.e., the highest oscillation amplitudes give the lowest asymptotic times, and they draw well determined regions. In fact, we can distinguish between the region of points with high oscillations amplitude (low asymptotic times) and the one with low oscillations amplitude (high asymptotic times).  As a result, when the parameters are fixed inside those regions some predictions can be possible. On the contrary, on the boundaries of the regions the points are well intermingled, showing fractal behavior. Here, the possibility to predict the asymptotic times when the parameters are fixed on the regions boundaries is impossible. This asserts our previous statement that the $\alpha$ parameter can change greatly the asymptotic time and the dynamics of the system. Therefore, it could be used instead of a long and complicated damping term since it can mimics the same dynamics.

\section{Analysis and results in the overdamped case}\label{sec:5}

In the last part of our study, we add the effects of large values of the damping parameter $\mu$ on the asymptotic behavior of the Duffing oscillator. In the non-fractional case, high values of $\mu$ means that the dynamics of the system is quite trivial and all trajectories fall down into one of the fixed points. However, in the fractional case, the fractional derivative multiplied by a large damping parameter can play a relevant role since the dynamics of the system becomes not so trivial in different ranges of $\alpha$. Previous work reveals the effects of  $\alpha$ for the overdamped case in the Helmholtz oscillator \cite{Coccolo_fr}. Here, we study the $\alpha$ parameter effects on the oscillations amplitude and on the asymptotic times for an overdamped regime in which the value of the damping parameter is fixed to $\mu=0.8$. Firstly and due to the large dissipation, in all the simulations the asymptotic times decrease, so we have fixed the time of integration at $t=300$, which for our purposes is more than enough to reach the steady state.

\subsection{Overdamped case with $\omega=1$}

\begin{figure}[htbp]
  \centering
   \includegraphics[width=16.0cm,clip=true]{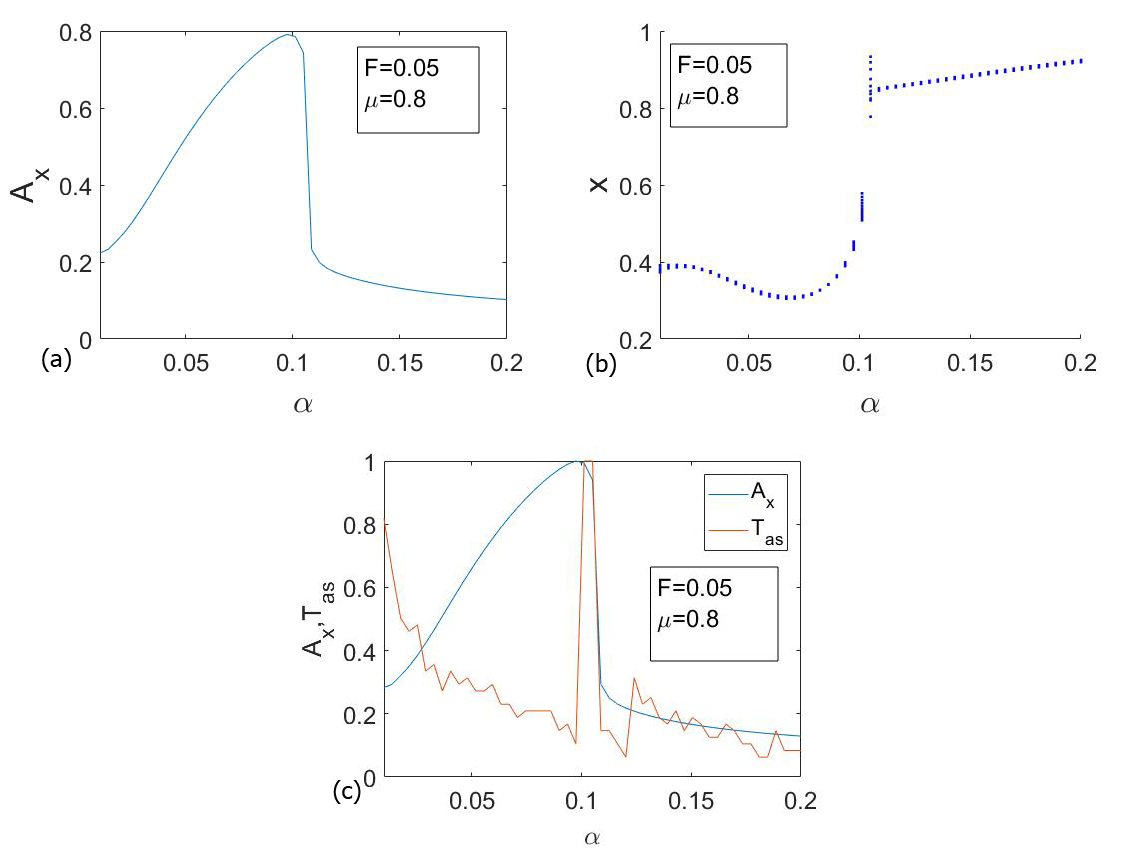}
   \caption{Panel (a) shows the amplitude of the oscillations for $F=0.05$ and $\mu=0.8$. This figure is compared with a diagram of the asymptotic behaviors (panel~(b)), in which it is possible to see that the peak is related with a change in the asymptotic behavior of the system. Panel (c) shows both the amplitude of the oscillations and the asymptotic time normalized in the same plot. We can see that the peak in the $T_{as}$ is related with the peak in the amplitude, and the change in the asymptotic behavior (see panel~(b)).}
\label{fig:9}
\end{figure}

Now, we analyze the effects of the dissipation for values of the forcing amplitude of $F=0.05,~F=0.07,~F=0.09$ and $F=0.1$ by fixing the forcing frequency to be $\omega=1$, as in the underdamped case in order to compare both situations. Here, we can distinguish interesting phenomena which do not take place in the non-fractional case.\\
In Fig.~\ref{fig:9}, we show the results for $F=0.05$. There, we can observe that the amplitude of the oscillations reaches a maximum for $\alpha=0.1$ (see Fig.~\ref{fig:9}(a)), which is related with a change in the asymptotic behavior of the system, as shown in Fig.~\ref{fig:9}(b). Besides, it may be noticed that the amplitude of the peak, and thus the change in the asymptotic behavior, corresponds with a maximum in the asymptotic time, $T_{as}=300$.

\begin{figure}[htbp]
  \centering
   \includegraphics[width=16.0cm,clip=true]{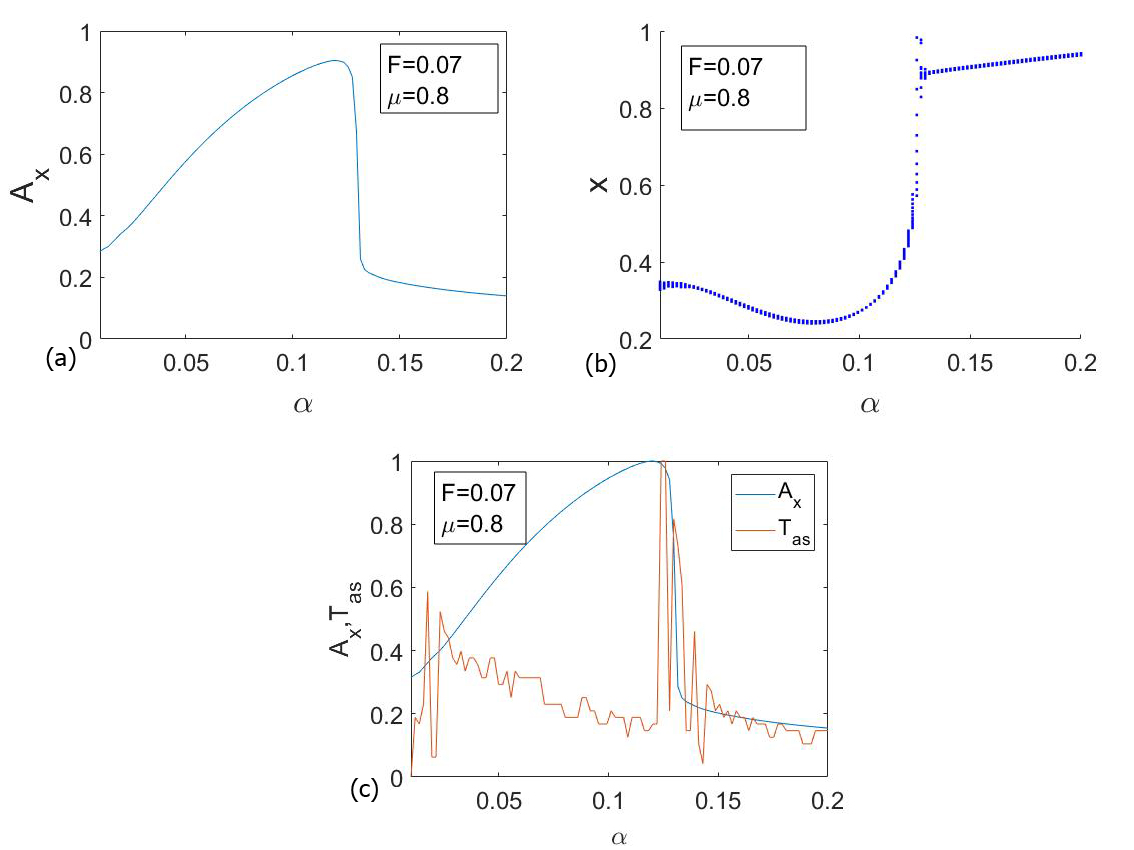}
   \caption{Panel~(a) shows the amplitude of the oscillations for $F=0.07$ and $\mu=0.8$. This figure is compared with a diagram of the asymptotic behaviors (panel~(b)), in which it is possible to see that the amplitude peak is related with a change in the asymptotic behavior of the system. Panel~(c) shows both the amplitude of the oscillations and the asymptotic time normalized in the same plot. We can see that the peaks in the $T_{as}$ are related with the peak in the amplitude, and the change in the asymptotic behavior, (see panel~(b)).}
\label{fig:10}
\end{figure}

\begin{figure}[htbp]
  \centering
   \includegraphics[width=16.0cm,clip=true]{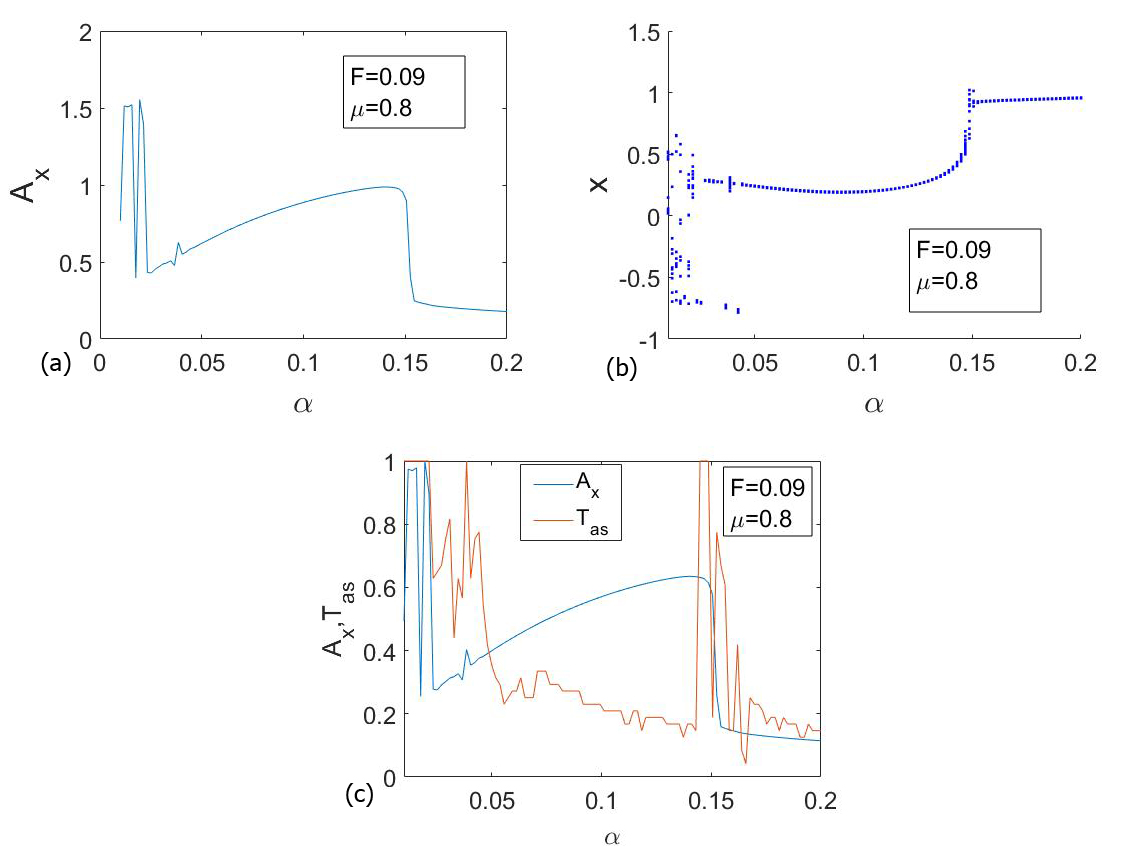}
   \caption{Panel~(a) shows the amplitude of the oscillations for $F=0.09$ and $\mu=0.8$. This figure is compared with a diagram of the asymptotic behaviors (panel~(b)), in which it is possible to see that the amplitude peaks are related with a change in the asymptotic behavior of the system. Moreover, it is possible to recognize a chaotic region  for small values of the $\alpha$ parameter. Panel~(c) shows both the amplitude of the oscillations and the asymptotic time normalized in the same plot. We can see that the peaks in the $T_{as}$ are related with the peak in the amplitude, and the change in the asymptotic behavior, (see panel~(b)). The chaotic region is related with the maximum of the asymptotic time.}
\label{fig:11}
\end{figure}

In addition, Fig.~\ref{fig:10} shows the same plots as in the previous case but for $F=0.07$. We can observe that any significant differences can be found except for the fact that the amplitude peak appears for higher value of the $\alpha$ parameter, $\alpha\simeq0.13$. However,  in Fig.~\ref{fig:11}, in which $F=0.09$, a chaotic region appears for small $\alpha$ values ($\alpha<0.0215$), that corresponds with a maximum in $T_{as}$. After some numerical simulations, we find out that this chaotic region starts to appear in the asymptotic behaviors diagram for the forcing amplitude value $F=0.08$, but it extends only a little bit beyond $\alpha=0.01$. Other simulations, not displayed here to avoid redundancy, show that the higher the value of the forcing amplitude the wider the chaotic region along the $\alpha$ axis.

It is important to stress out that in all figures in which asymptotic times are shown, Figs.~\ref{fig:9}(c),~\ref{fig:10}(c) and~\ref{fig:11}(c), two well-defined regions on the $\alpha$ axis show up, the first one before the amplitude peak and the other beyond it. We can observe that in the first region that the asymptotic times follow a decreasing trend,  which actually  continues after the amplitude peak. This behavior is different from the underdamped case. Figure~\ref{fig:11}(c), for $\alpha<0.0215$, does not follow the same trend due to the existence of the chaotic region but beyond that region the decreasing trend is recognizable.
Therefore, we can say that the effect of a large dissipation is to stabilize the trend of the asymptotic time. So it seems that in the overdamped case it would be easier to predict the behavior of the asymptotic times, making it easier to use the fractional derivative to model materials with complicated dissipation.

\begin{figure}[htbp]
  \centering
   \includegraphics[width=16.0cm,clip=true]{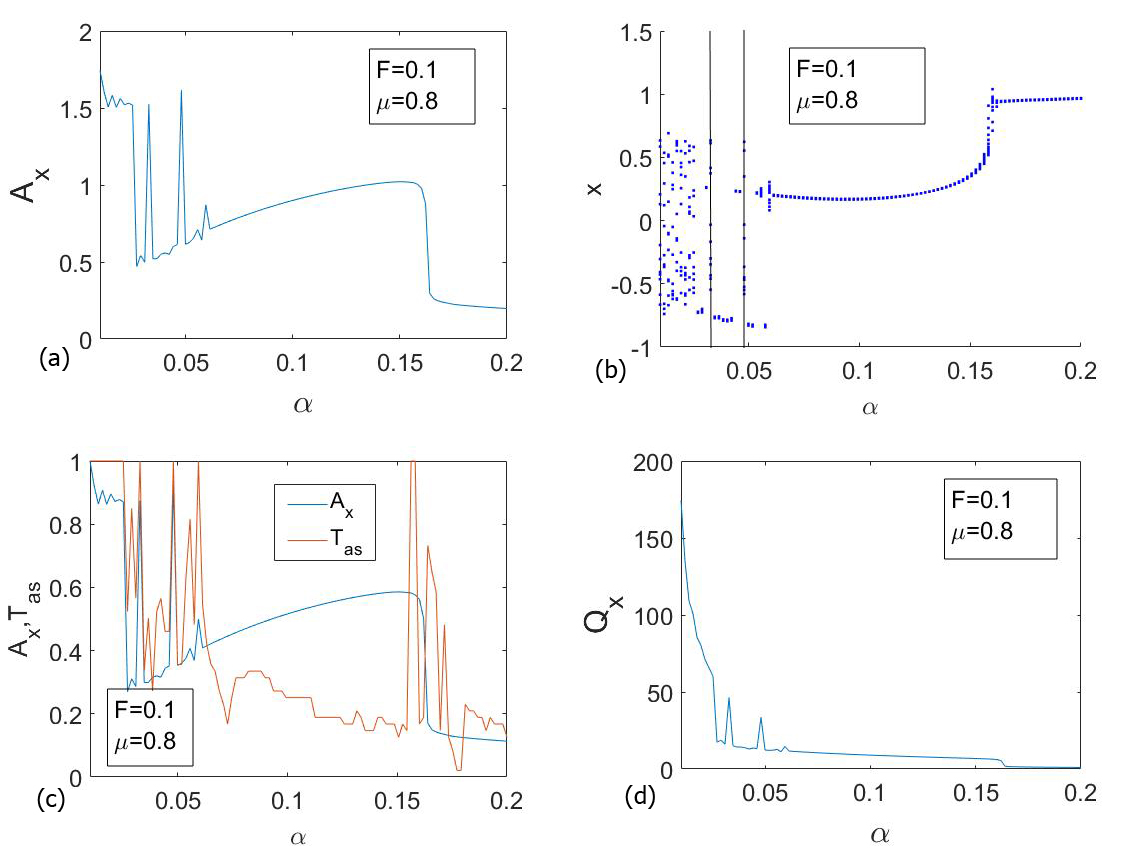}
   \caption{Panel~(a) shows the amplitude of the oscillations for $F=0.1$ and $\mu=0.8$. This figure is compared with a diagram of the asymptotic behaviors (panel~(b)), in which it is possible to see that the amplitude peak related with a change in the asymptotic behavior of the system happen at $\alpha\simeq0.152$. Then, there are two peaks at $\alpha=0.0328$ and $\alpha=0.048$ correlated with chaotic behaviors of the system, marked by the two vertical black lines in panel~(b). Again, it is possible to appreciate that there is a chaotic region for small values of the $\alpha$ parameter. Panel~(c) shows the amplitude of the oscillations and the asymptotic time normalized in the same plot. We can see that the peaks in the $T_{as}$ are related with the peak in the amplitude, (panel~(a)), and the change in the asymptotic behavior, (see panel~(b)). The chaotic region is related with the maximum of the asymptotic time. Finally, in panel~(d) we present the $Q-$factor that show that a sort of resonance phenomenon has been triggered.}
\label{fig:12}
\end{figure}

Finally, in Fig.~\ref{fig:12}, we analyze the case for $F=0.1$. Here, we can note that, again, beyond the chaotic region and before and after the amplitude peak at $\alpha\simeq0.15$, the asymptotic times follow a decreasing trend.  Besides, in Fig.~\ref{fig:12}(a), we can see that there are two specific amplitude peaks that are related to orbits that reach both wells at $\alpha=0.0328$ and $\alpha=0.048$, respectively. These two peaks correspond to the two vertical lines plotted in Fig.~\ref{fig:12}(b), that mark two chaotic behaviors of the system.

The related asymptotic time is maximum, as shown in Fig.~\ref{fig:12}(c), because the behavior of the system is chaotic. On the other hand, for values of the $\alpha$ parameter just before or beyond those lines, the orbits stay in one well.  In this case, it is useful to plot also the $Q-$factor, Fig.~\ref{fig:12}(d). In fact, the two peaks are recognizable and they are qualitatively different from what we have shown in Fig.~\ref{fig:5}.  This makes us to think that a sort of resonance phenomenon can be triggered due to the influence of a large dissipation for this threshold value of the forcing amplitude and these two values of the $\alpha$ parameter, $\alpha=0.0328$ and $\alpha=0.048$. In other words, the system needs a certain amount of forcing amplitude to give the birth of those peaks for those values of $\alpha$. Besides, it also needs a strong dissipation to suppress the interwell orbits for the others values of $\alpha$. The underdamped regime cannot achieve this resonant-like behavior since it does not suppress the interwell orbits for these values of $\alpha$. Then, we have analyzed the case of higher forcing amplitude, $F=0.11$ and $F=0.12$, and find out that similar peaks appear but for other values of the parameter $\alpha$.

\FloatBarrier

\FloatBarrier

\FloatBarrier

\FloatBarrier

 \begin{figure}[htbp]
  \centering
   \includegraphics[width=16.0cm,clip=true]{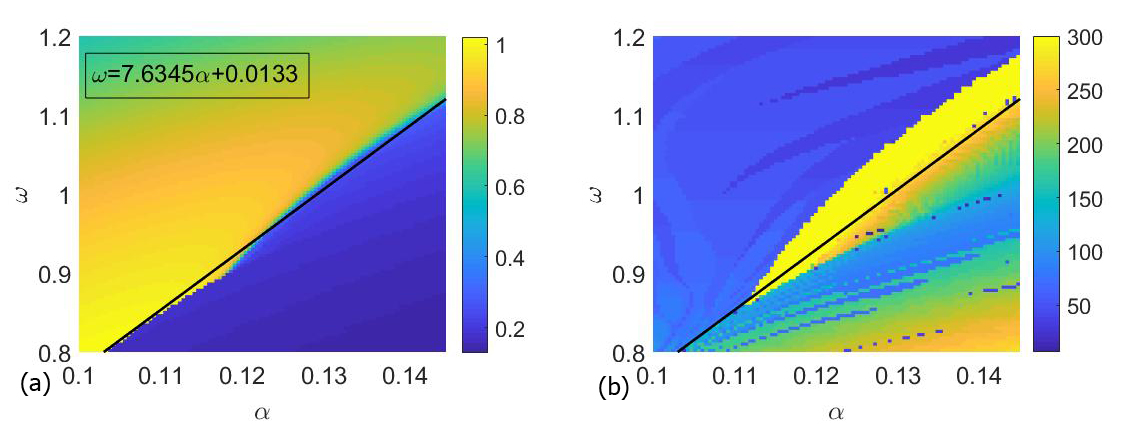}
   \caption{The left panel~(a) shows,  for $F=0.07$, the amplitude parameter set while the right panel (b) shows  the asymptotic time parameter set in function of the parameters $\alpha$ and $\omega$. In panel~(a) is, also, plotted the solid black line that separates the high amplitude region from the low amplitude one according to the equation $\omega=7.6345 \alpha +0.0133$. Here, we can appreciate that the highest amplitudes in the right figure are almost related to the lowest asymptotic times. In fact, there are also regions in which the asymptotic times can show a complicated behavior even if we are studying the overdamped case. The same black line is plotted in panel~(b), but its division of the asymptotic time parameter set is not as accurate as in the amplitude parameter set.}
\label{fig:13}
\end{figure}
\FloatBarrier

 \begin{figure}[htbp]
  \centering
   \includegraphics[width=14.5cm,clip=true]{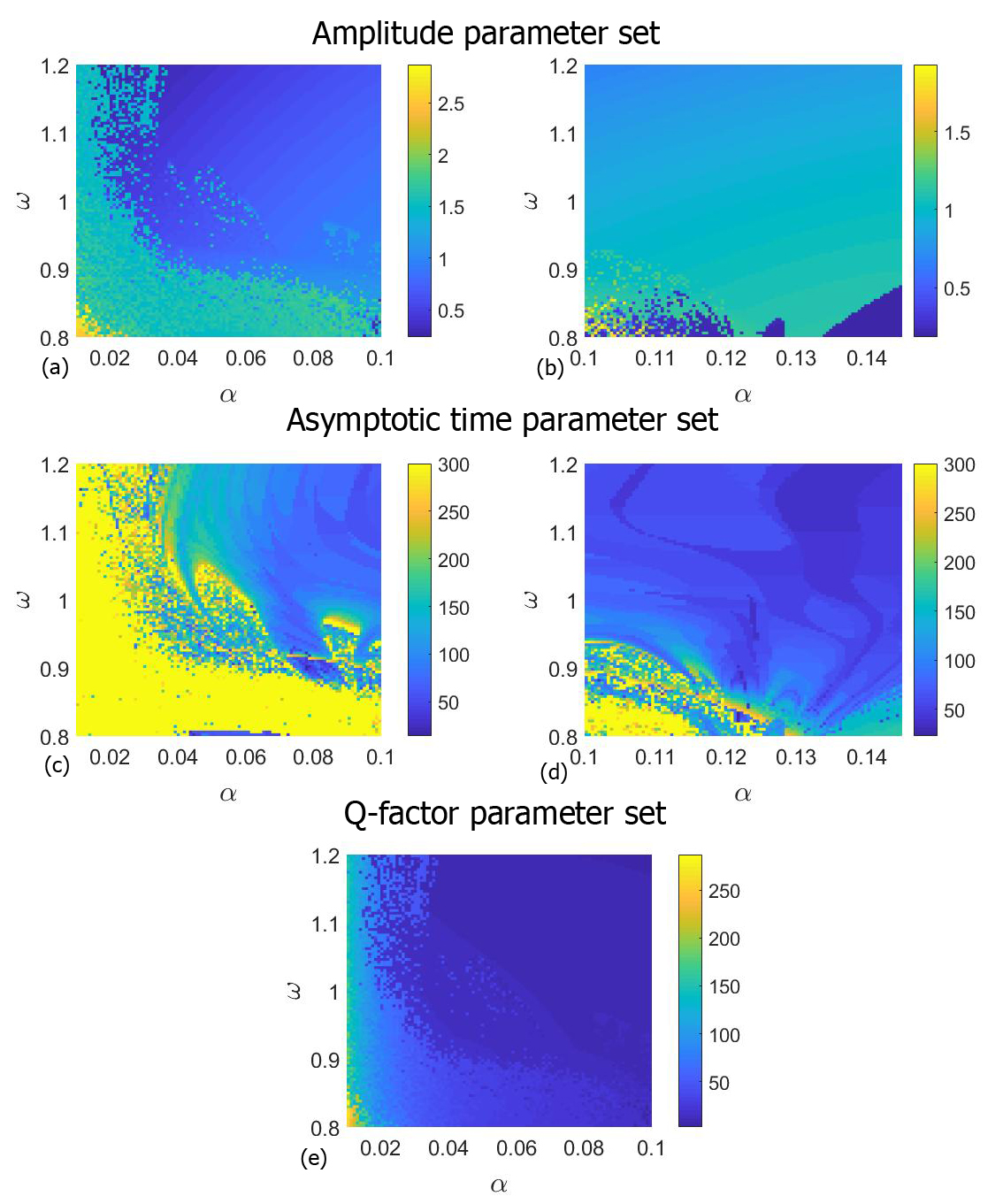}
   \caption{Panels (a) and (b) show the oscillations amplitude parameter set. Panels (c) and (d) show the asymptotic time parameter set and panel~(e) shows the $Q-$factor parameter set. All of them are plotted in function of the parameters $\alpha$ and $\omega$ and for a forcing amplitude $F=0.1$. Panels (a) and (e) show that the higher amplitudes are concentrated at small $\alpha$ and $\omega$ values. In all the figures we can appreciate that there are regions that show fractality. This means that, for this value of the forcing, the oscillation amplitudes and the asymptotic times can show a complicated behavior even in the overdamped case. }
\label{fig:14}
\end{figure}
\FloatBarrier

\subsection{Overdamped case for $0.8\leq\omega\leq1.2$}

Now, we analyze the overdamped case for the forcing frequency values $0.8\leq\omega\leq1.2$, as in the underdamped case, in order to investigate the impact of the fractional parameter for values of the frequency near $\omega=1$ in presence of large dissipation. For this purpose, we plot the oscillations amplitude and the asymptotic times parameter set as a function of both the fractional parameter and the forcing frequency.  In Fig.~\ref{fig:13}, we show the parameter sets for $F=0.07$, which is a simple case with no chaotic regions nor resonance-like peaks. Thus, as it is possible to anticipate by taking a look of Figs.~\ref{fig:10}, nothing particular is expected in the amplitude parameter set. In Fig.~\ref{fig:13}(a), the only remarkable thing is that the jump from a high to a small amplitude occurs for different $\alpha$ values insofar $\omega$ changes. Moreover, the $\omega$  value responsible for the change in amplitude seems, to be directly proportional to the $\alpha$ parameter approximately according to the function $\omega=7.6354\alpha+0.0133$. It is important to stress out that this is not a linear fit, but an estimation we have made by taking the two maximum points on the edge between the two region at $\omega=0.8, \alpha=0.103$ and $\omega=1.12, \alpha=0.145$ and connect them trough a line.

Being able to calculate the proportionality of the two parameters, we can have the possibility to control the outcome of the oscillations amplitude by fixing a value of the forcing frequency and a value of the $\alpha$ parameter. It is not the same for the asymptotic times, (see Fig.~\ref{fig:13}(b)), that follow a similar but also different path with respect to the amplitude parameter set. In fact, Fig.~\ref{fig:13}(b) shows a more complicated geometry. Although the left upper part of the figure resembles Fig.~\ref{fig:13}(a), the regions division does not follow the line sketched in the amplitude parameter set. Moreover, the right lower part is different and shows a more intermingled framework.  This means that the asymptotic time regions can develop a complicated structure even in the overdamped case. So, according to Fig.~\ref{fig:13}(a), for a given $\omega$, it is possible to fix a certain value of $\alpha$ to obtain a high or a low oscillations amplitude. For example for $\omega=0.9$, we can fix $0.1<\alpha<0.11$ to obtain higher oscillations amplitudes or we can fix $\alpha>0.12$ to obtain low oscillations amplitudes. On the contrary, for the asymptotic times it is not as easy as for the oscillations amplitudes, although it is easier than in the underdamped regime.

Then, Figs.~\ref{fig:14}(a-e) have been obtained for $F=0.1$, value for which in Fig.~\ref{fig:12} resonance-like peaks and a chaotic region show up. The Figs.~\ref{fig:14}(a) and~\ref{fig:14}(b) show that the higher oscillations amplitudes are concentrated at small $\alpha$ and $\omega$ values, exactly the opposite occurs for the asymptotic time in Figs.~\ref{fig:14}(c) and~\ref{fig:14}(d). It can be seen that the regions formed in the parameter sets show fractality and complicated behaviors in both amplitudes and asymptotic times parametric sets. This means that, again, the fractional derivative parameter can be used to explain complicated behaviors of different kind of material, although the possibility to predict them is reduced with respect to the $F=0.07$ case. Finally, Fig.~\ref{fig:14}(e) shows the $Q-$factor parameter set, related with the resonance-like phenomenon. Here, we can see that the region of interwell oscillation amplitudes of Fig.~\ref{fig:14}(a) has a higher $Q-$factor, suggesting the aforementioned resonance-like phenomenon.

\section{Conclusions and discussion}\label{sec:conclusions}
We have studied the fractional Duffing oscillator with a damping fractional term, using the Gr\"unwald-Letnikov integrator. The results of this paper can be summarized by saying that the fractional parameter $\alpha$ plays a relevant role in the changes of the system amplitudes and has a huge impact on its asymptotic time, which is the time for which the system reaches its steady state. In the first part of this work, which has been focused on the underdamped case, we have shown that the dynamics of the system change insofar $\alpha$ is varying and how some peaks in the amplitude plots appear. Besides, for $F=0.07$, a chaotic region emerges for small values of the $\alpha$ parameter. In the second part, also in the underdamped situation, we have shown how these changes in the dynamics provoke big variations in the asymptotic time of the system. In particular, the asymptotic times fall to the smallest values when the amplitude reaches the maximum.

Finally, we have extended our previous study to the overdamped case in which we have found regions of chaotic motions for different values of the parameter $\alpha$ that do not appear in the non-fractional case. Moreover, we have seen the effect of a large dissipation on the asymptotic times. In fact, a decreasing trend appears as the fractional parameter grows. On the contrary, in the overdamped case, the combination of a large dissipation and a certain threshold of the amplitude forcing $F=0.1$, triggers peaks like in a resonance phenomenon. So, it seems that the fractional parameter can trigger a resonance-like phenomenon when the dissipation is large enough. In summary, we think that this work can be useful for potential applications as in the study of the materials behaviors that cannot be described by a simple damping term, among others.

\section{Acknowledgment}

This work has been supported by the Spanish State Research Agency (AEI) and the European Regional Development Fund (ERDF, EU) under Project No.~PID2019-105554GB-I00.

\end{document}